\documentclass[preprint,showpacs,showkeys,preprintnumbers,amsmath]{revtex4}
\usepackage{amsmath}
\usepackage{graphicx}
\usepackage{dcolumn}
\usepackage{bm}

\begin{document}

\title{Self-trapping at the liquid-vapor critical point}

\author{Bruce N. Miller\email{B.Miller@tcu.edu}}
\affiliation{Department of Physics, Texas Christian University, \\
Fort Worth, Texas 76129}
\author{Terrence L. Reese\email{Terrence_Reese@cxs.subr.edu}}
\affiliation{Department of Physics, Southern University and A\&M
College,
\\Baton Rouge, Louisiana 70813}
\date{\today}



\begin{abstract}

Experiments suggest that localization via self-trapping plays a
central role in the behavior of equilibrated low mass particles in
both liquids and in supercritical fluids. In the latter case, the
behavior is dominated by the liquid-vapor critical point which is
difficult to probe, both experimentally and theoretically. Here,
for the first time, we present the results of path-integral
computations of the characteristics of a self-trapped particle at
the critical point of a Lennard-Jones fluid for a positive
particle-atom scattering length. We investigate the influence of
the range of the particle-atom interaction on trapping properties,
and the pick-off decay rate for the case where the particle is
ortho-positronium.

\end{abstract}

\maketitle

The system consisting of a massive particle equilibrated in a host
fluid, known to physicists and chemists as Brownian Motion, has
played a seminal role in the development of Statistical Physics
\cite{Ei,MS}. However, the opposite regime of an equilibrated low
mass particle is equally challenging and manifests a richer set of
behaviors \cite{HerRMP}. Except at very high temperatures, quantum
mechanics is required to model the low mass particle. Possible
quantum particle (qp) candidates are an electron, positron, or
positronium atom, while the host can either be a dense gas or
liquid below the critical temperature, or a supercritical fluid
above it. Experimental measurements of the properties of a low
mass particle equilibrated in a fluid strongly suggest that it can
induce a local, mesoscopic, deformation in the fluid in which the
qp becomes self-trapped, or localized. \cite{PosPhys,HerRMP} Since
the qp has a long deBroglie wavelength, intuitively we anticipate
that it simultaneously interacts with a large group of atoms or
molecules in the host fluid forming a mesoscopic region of altered
fluid density. Depending on whether the effective qp-atom
interaction is attractive or repulsive, the local density of the
host is either augmented or suppressed near the qp, resulting in
the formation of either a ''microdroplet'' or ''microbubble''. The
intuitive picture is completed by imagining that the qp occupies
the ground state of the potential well induced by the formation of
the density inhomogeneity, i.e. the droplet or bubble, thus
stabilizing the deformation.

The positron-atom interaction is characterized by a negative
scattering length so the experimental manifestation of
self-trapping is a decrease in the positron lifetime due to the
increase in the local electron density resulting from the
formation of a droplet. The reverse is true for ortho-positronium.
Angular momentum conservation eliminates the two photon decay
process for the triplet state. As a result of the long natural
lifetime of o-Ps in the vacuum, about 142 ns \cite{PosPhys}, its
positron can annihilate more readily with an atomic electron of
the host. A consequence of the fermionic repulsion between the
electron in o-Ps and the electrons of the host fluid is that this
''pick-off'' annihilation rate is reduced when self-trapping
occurs, demonstrating bubble formation \cite{PosPhys}. Depending
on the choice of the host, the electron-atom scattering length can
have either sign, so each behavior is possible. While the most
dramatic manifestation of self-trapping of the positron or
positronium is a significant deviation from linearity in Arrhenius
plots of the decay rate versus average density, the signature of
electron self-trapping is a change in mobility \cite{BorgSan}.

In addition to the liquid state \cite{Dutta}, self-trapping occurs
in a broad, super-critical region of density, $\rho$, and
temperature, $T$, surrounding the liquid-vapor critical point
$(\rho _{c},T_{c})$ \cite{PosPhys,IK}. Since the isothermal
compressibility diverges at the critical point, this is not
surprising: The qp can more easily alter the local density in this
region of temperature and density. However, as a result of the
large density gradient induced by the earth's gravitational field,
there are few reliable experimental studies of self-trapping close
to the critical point \cite{Sharma84}. Thus, although it dominates
the self-trapping regime above $T_{c}$, the effect of close
proximity to the critical point on self-trapping is not generally
known \cite{Sharma84}.

The theory of self-trapping has evolved through different stages:
In the earliest models, the qp occupies the ground state of a
spherical step potential which is assumed to be proportional to
the local fluid density \cite{Fer}. In modelling the qp-host
interaction the atomic nature of the host is ignored, and it is
simply represented by a type of jellium. By minimizing the free
energy of the qp-fluid system, it is possible to show that the
deformation is stable in a bounded region of temperature and
density near the critical point \cite{Fer}. Later versions of mean
field theory (MFT) permitted a continuous density profile
\cite{IK,SZ} and took into account the atomic nature of the host
at an intermediate level \cite{Niem80}. An interesting improvement
was obtained by employing the Percus-Yevick equation to include
the effect of qp-atom and atom-atom correlations in the mean field
formalism \cite{YanTsai}. An advantage of MFT is that computations
are reduced to numerically integrating a pair of coupled ordinary
differential equations. However, in practice, they have only
proved useful for fluids at low temperatures
\cite{SZ,IK,YanTsai,RM89}. This problem may arise because mean
field theories only include a single bound state for the qp. More
microscopically complete models have evolved during the last few
decades based on the Feynman/Kac path integral \cite{FeySM} which
overcome the major shortcomings of the earlier work. These account
for the details of the qp-atom interaction and implicitly take
into account local fluctuations in the disturbed fluid and the
state of the qp. Two approaches have been employed, one based on
an approximate analytic model requiring an educated choice of
closure \cite{ChanSR,CM94}, and the more direct alternative using
Monte Carlo algorithms \cite{CokBer,RM89}. While the latter avoids
approximation, the computational cost is greater.

Here we employ path integral Monte Carlo (PIMC) to investigate
self-trapping at the liquid-vapor critical point. We take
advantage of recent improvements in fluid equilibrium theory which
give accurate critical point parameters for the Lennard-Jones
(6,12) potential.\cite{Wil,White} We model the qp-atom interaction
with a hard-sphere potential and study the dependence of the
physical and statistical properties of self-trapping on the hard
sphere diameter, $R_{hs}$. In particular, we investigate the
dependence on $R_{hs}$ of both the spreading of the qp
wavefunction and its influence on the local deformation of the
fluid. Since a repulsive qp-atom interaction is representative of
positronium \cite{PosPhys}, we also estimate the pick-off
annihilation rate for a simplified model of the atomic charge
distribution.

Although no experimental measurements of the pick-off decay rate
have been carried out precisely at the critical point, for the
case of Xenon they exist for two supercritical temperatures,
$300K$ and $340K$, over a large density range and strongly suggest
the existence of the Ps self-trapped state \cite{Tuom}. In
previous work \cite{RM01} we have used PIMC to carry out
simulations of Ps in Xenon at these temperatures, so it is natural
to select the critical point of Xenon as a test case and a mass of
$2m_{e}$ for the qp. The truncated version of the Lennard-Jones
6-12 potential was chosen to represent the inter-atomic potential.
Wilding has established numerically accurate connections between
the L-J distance and energy parameters ($\sigma $ and $\varepsilon
$) and the critical temperature and density ($T_{c}$ and $\rho
_{c}$), namely $\rho _{c}\sigma ^{3}=0.3197 $ and
$kT_{c}/\varepsilon =1.1876$ \cite{Wil}. The experimental values
of the critical density and temperature of Xenon are $T_{c}=289K$
and $\rho_{c}=5.299\times 10^{-3}$ atoms/\AA$^{3}$ yielding
$\sigma =3.92$\AA\ and $\varepsilon /k=243.5K$.

The computational simulation of the qp-fluid system using a
complete quantum mechanical description would be intractable; thus
approximate representations of the system are required. Except at
very low temperatures, the translational degrees of freedom of an
atomic fluid can be approximated with classical mechanics. Thus
the qp-fluid system is well represented by a hybrid
classical-quantum Hamiltonian:
\begin{equation}
H=\sum_{1}^{N}\mathbf{P}_{j}^{2}/2M+\sum_{N\geq j>k\geq 1}U\left(
\left| \mathbf{R}_{j}-\mathbf{R}_{k}\right| \right) +H_{qp},
 ~~H_{qp}=-\frac{\hbar ^{2}}{2m}\Delta +\sum_{1}^{N}V\left( \left|
\mathbf{r}-\mathbf{R}_{j}\right| \right).
\end{equation}
Here $\mathbf{P}_{j}$ and $\mathbf{R}_{j}$ are the momenta and
positions of the $N$ host atoms, $\mathbf{r}$ is the position of
the qp, while $U$ and $V$ are the pairwise additive atom-atom and
qp-atom interaction potentials. This is known as the adiabatic
formulation and results in a hybrid partition function,
\cite{RM01}
\begin{equation}
Z=\int d\underline{\mathbf{R}}e^{-\beta U(\underline{\mathbf{R}})}
\int d\mathbf{r} \langle \mathbf{r}| e^{-\beta H_{qp}}| \mathbf{r}
\rangle /(N!\Lambda^{3N}),
\end{equation}
where here $\underline{\mathbf{R}}$ represents
$\{\mathbf{R}_{j}\}$, the complete $3N$ dimensional configuration
space of the atoms with classical potential energy
$U(\underline{R})$, and $\Lambda$ is the atomic thermal
wavelength. Thus, in principle, the quantum statistical average of
the physical operator $O$ can be computed from
\begin{equation}
\langle O\rangle = \int d\underline{\mathbf{R}}e^{-\beta U(
\underline{\mathbf{R}}) } \int d\mathbf{r} \langle \mathbf{r}|
e^{-\beta H_{qp}}O| \mathbf{r}\rangle /(N!\Lambda^{3N}Z).
\end{equation}
To obtain a formulation which is useful for computation, we follow
the Feynman-Kac path integral construction \cite{FeySM}. First, by
applying the Trotter formula to $Z$, we can express the trace over
$\mathbf{r}$ as a sum over discretized paths of $P$ steps. For
sufficiently large $P$ the kinetic and potential energy operators
approximately commute in each step, yielding the following
expression for the partition function;
\begin{equation}
Z_{P}=\int d\underline{\mathbf{R}} e^{-\beta
U(\underline{\mathbf{R}})}[ \prod_{0}^{P-1}\int d\mathbf{r}_{i}
e^{-P| \mathbf{r}_{i}-\mathbf{r}_{i+1}| ^{2}/2\lambda ^{2}}
\rho(\mathbf{r}_{i},\mathbf{r}_{i+1},\underline{\mathbf{R}},\beta
/P)],
\end{equation}
where $\lambda$ is the qp thermal wavelength. If we take $\rho$ to
be $exp(-\beta V/P)$, equation (4) is known as the primitive
approximation. Thus, in the discretized Feynman-Kac path integral
\cite{FeySM}, the qp is represented by a closed chain of $P$
classical "pseudo-particles", or "slices" in imaginary time
$t_{i}=(i/P)\beta\hbar$, with harmonically coupled nearest
neighbors. Each pseudo-particle in the chain interacts with each
atom through the potential $V/P$.The chain is equilibrated in a
fluid at the augmented temperature $PT$ \cite{CokBer,Sprik}. In
the limit $P\rightarrow \infty$ the correspondence is exact.The
spread of the chain corresponds to the uncertainty in position of
the qp. The equivalence allows Monte Carlo methods developed for
classical systems to be used to compute quantum mechanical
equilibrium properties. Path integral Monte Carlo has been used
successfully by many groups, including our own, to compute the
equilibrium properties of quantum systems \cite{QMC}.

To insure convergence, while maintaining a constant temperature
and average atomic density, the system size, the number of
chain-particles $P$, and the number of statistical samples are
increased until there are no significant changes in the calculated
equilibrium properties. The algorithm we employed here depends on
five parameters, the density, temperature, $R_{hs}$, the number of
fluid atoms, $N$, and $P$. Convergence was improved by using an
image potential to smooth out the singularity of the hard sphere,
qp-atom interaction \cite{WhitKal,RM01}. A complete description of
this method is given in our previous publications (see \cite{RM01}
and references cited within). For this initial study we selected
three values of $R_{hs}$; $0.5$\AA\ , $5.0$\AA\, and $9.5$\AA\
while $P$ was chosen to be $2000$ as in our previous computations
\cite{RM01}. Since most of the cpu time was expended on
repositioning the atoms, the large value of $P$ did not pose a
problem and insured convergence. To minimize edge effects, the
length of the box in which the simulations were carried out was
set to three times the qp thermal wavelength.

Important characteristics of the qp are the root mean square
displacement ($D(t)$) between two particles along the chain
separated by imaginary time $t$, and the density of chain
particles a distance $r$ from its center of mass ($\rho_{cm}(r)$).
$D$ and $\rho_{cm}$ are both direct measures of self-trapping. In
an extended state the displacement between particles separated by
$t$ increases with $t$, reaching a maximum at $t=P/2$, according
to a unimodal parabolic distribution. However, in a self-trapped
state the chain is highly confined and thus, except for values of
$t$ near the end points at $0$ and $P$, $D(t)$ is roughly
constant. The function $\rho_{cm}(r)$ represents the mean
probability density of the qp wavefunction, and becomes more
peaked around the chain com as the chain becomes confined within
its self-trapped bubble. A detailed discussion of these quantities
and how they are computed can be found in our previous paper on
self-trapping of Ps in Xenon \cite{RM01}.

\begin{figure}[tbp]
\centerline{\includegraphics{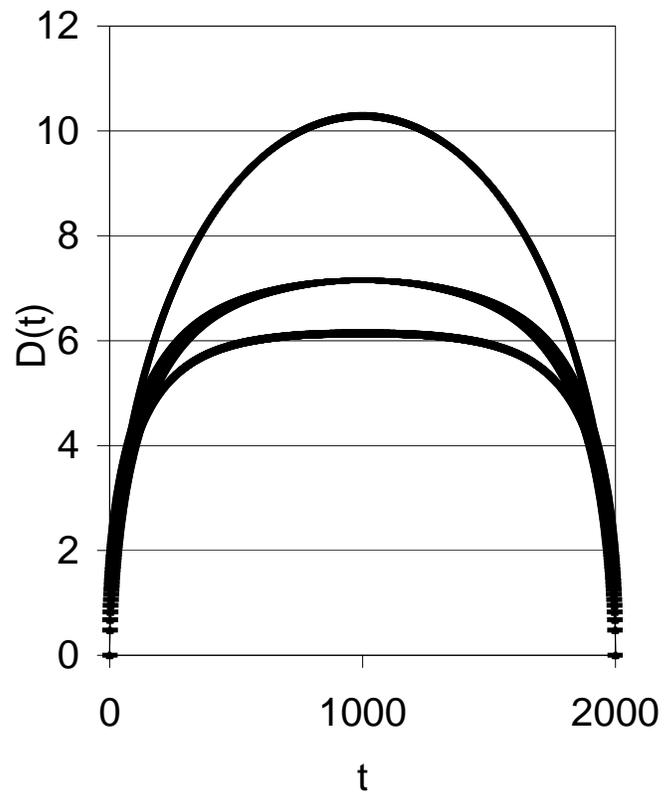}}
\caption{Plot of
$D(t)$(Angstroms) versus $t$ for the $R_{hs}$ values $0.5$\AA~
(upper curve), $5.0$\AA~ (middle) and $9.5$\AA~ (bottom).}
\label{Dpre}
\end{figure}

\begin{figure}[tbp]
\centerline{\includegraphics{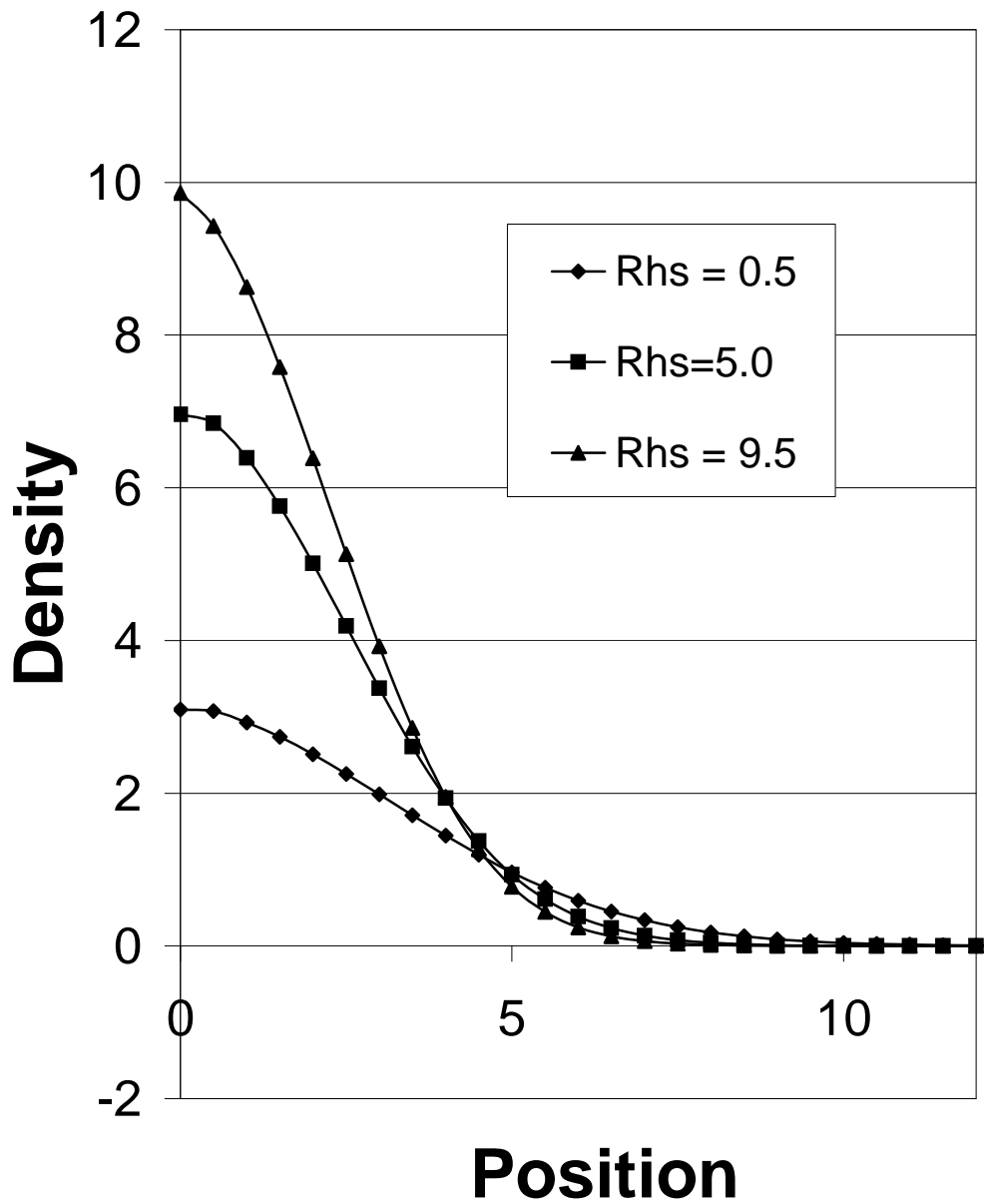}} \caption{Density of chain
particles as a function of distance from the chain centroid,
$\rho_{cm}(r)$, for the smallest ($0.5$\AA), average ($5.0$\AA),
and largest ($9.5$\AA) values of $R_{hs}$.} \label{rhocmpre}
\end{figure}

Figure~\ref{Dpre} is a plot of $D$ versus $t$ for the three
$R_{hs}$ values considered here. As $R_{hs}$ increases, the figure
shows that the shape of the plots changes from parabolic,
corresponding to a nearly free particle, to one that is
essentially constant in the central region, indicating that the
chain becomes more confined \cite{Sprik,CokBer}. Increasing the
value of $R_{hs}$ expels fluid atoms from the vicinity of the
chain particles into regions which don't overlap with them. At the
same time the pressure applied by the fluid atoms compresses the
chain.  Since there are no repulsive forces between the
pseudo-particles comprising the chain, their density near the
chain com can become quite large. This can be seen in
Figure~\ref{rhocmpre}, a plot of $\rho_{cm}$ versus distance from
the com, for the same $R_{hs}$ values shown in Figure~\ref{Dpre}.

\begin{figure}[tbp]
\centerline{\includegraphics {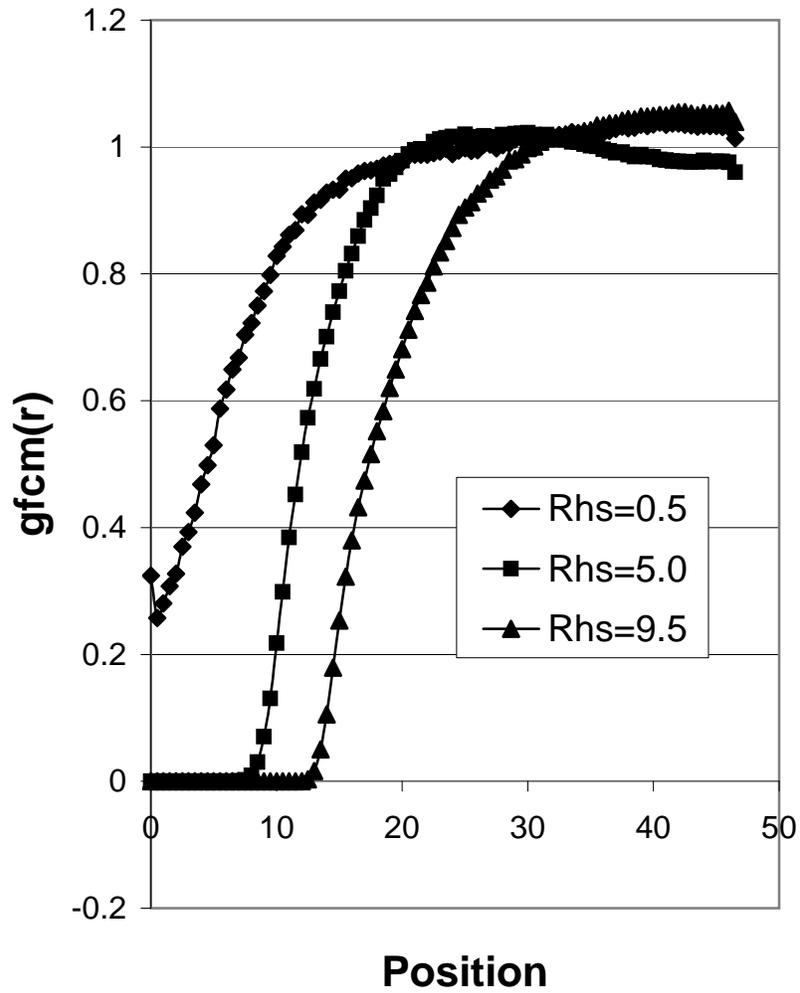}} \caption{Density of fluid
atoms from the chain center of mass, $g_{fcm}(r)$, versus position
for the smallest ($0.5$ \AA), average ($5.0$ \AA), and largest
($9.5$ \AA) values of $R_{hs}$.} \label{gfcmpre}
\end{figure}

Information concerning the local deformation of the fluid induced
by the qp is provided by the chain-fluid and com radial
distribution functions, $g_{fp}(r)$, $g_{fcm}(r)$, yielding,
respectively, the mean local density of fluid atoms a distance $r$
from a chain pseudo-particle, and the chain center of mass. As
usual, they are normalized to unity in the large $r$ limit. In
earlier work we showed that the former can be used to directly
determine the o-Ps pick-off decay rate \cite{RM01}, while the
latter provides direct evidence of self-trapping. In an extended
state fluid atoms are able to penetrate into the vicinity of the
chain com indicating that the qp is relatively spread out.  On the
other hand, in a completely self-trapped state, ground state
dominance prevails, the chain is folded upon itself within the
volume of the bubble, and fluid atoms are totally expelled from
this region. $g_{fcm}(r)$ can be used to directly compute the
average number of fluid atoms excluded from the vicinity of the
chain. Figure ~\ref{gfcmpre}, a plot of $g_{fcm}$ versus position,
indicates the strong exclusion of the fluid atoms from the
vicinity of the chain at the largest value of $R_{hs}$. The
significant value of $\rho_{cm}$ at the origin for the smallest
value of $R_{hs}$ provides evidence that the atoms can penetrate
the trapping region and suggests that the qp is in an extended
state. This is supported by the parabolic shape of $D(t)$. Hints
of slow oscillation on large scales, which we are unable to
resolve with the present system size, can be seen in
Figure~\ref{gfcmpre}. An open question is whether these are simply
finite size effects or represent a coupling between the
correlation length of the fluid and the thermal wavelength of the
qp. In our earlier work \cite{RM01} short range oscillations
occurred in $\rho_{cm}$ over distances on the order of $R_{hs}$ at
$\rho=2\rho_{c}$, indicating stacking of the fluid atoms in layers
around the qp com. We see from the figure that at the critical
point they are absent.

From $g_{fcm}$ we found that the volume of the excluded fluid
atoms from the trapping region is at least twice as great for
$R_{hs}=9.5$\AA~than for either $0.5$\AA~or $5.0$\AA. Since
increases in $R_{hs}$ will lead to decreases in the number of
fluid valence electrons near the qp available for annihilation,
for the case where the qp is o-Ps the pick-off decay rate is
expected to decrease with increasing $R_{hs}$. We computed the
decay rate at the three values of $R_{hs}$ considered above (see
table~\ref{t.1}). Taking ratios, we found an approximately
exponential decrease with $R_{hs}$ with a characteristic length of
$1.61$ \AA\ . In this model the Ps atom is treated as a composite
particle in its \textit{internal} hydrogenic ground state with an
exponentially decreasing wavefunction representing the spread of
the  positron from the Ps center of mass. This is justified by the
small ratio, $\approx4\times 10^{-3}$, of $kT_{c}$ to the
positronium excitation energy. For simplicity, the electron
density around the host fluid atoms is modelled as a delta
function. Thus the decay rate is dependent upon both the portion
of the positron ground state wavefunction that manages to leak
beyond $R_{hs}$ and the number of fluid atoms centers it overlaps.
The exponential decrease in the decay rate with increasing
$R_{hs}$ results from the decrease in both the density of fluid
atoms and the amplitude of the positron wavefunction in the  shell
surrounding the hard sphere surface.

In contrast with mean field models, the power of PIMC is that it
reveals the complete picture of both the distribution over qp
quantum states and the response of the fluid to the qp
''impurity''. In general we have found that compared with a liquid
\cite{Dutta} and our own PIMC computations at higher density
\cite{RM01}, the mesoscopic region in which the qp is localized is
much larger at the critical point. Moreover, the density profile
of the fluid has a different shape - the "walls" are not nearly as
steep and the structure is less sharply defined, i.e. it is
\textit{soft}. Our PIMC calculations for a supercritical fluid
with $\rho=\rho_{c}$, but at higher temperature, show similarities
with the critical point, but the behavior is less extreme
\cite{RM01}.

In future work we plan to employ PIMC to directly evaluate the
density of states. This will enable us to determine the angular
distribution of the annihilation photons from para-positronium,
which can be directly compared with experiment. To date this has
only been approached with a semi-empirical formulation of mean
field theory \cite{Dutta}. It will be interesting to see what
changes result from an ab initio computation which completely
includes the effects of correlated fluctuations.

\begin{table}
\caption{Natural log of the decay rate for three hard sphere
diameters.} \label{t.1}
\begin{center}
\begin{tabular}{lr}
$R_{hs}$ (\AA)  & Ln(Decay Rate)\\
0.5           &    -3.0  \\
5.0           &   -20.0  \\
9.5           &   -37.0  \\
\end{tabular}
\end{center}
\end{table}

\acknowledgements The authors benefitted from the support of the
Research Foundation and the Division of Information Services of
Texas Christian University. They are also grateful for the
computer expertise provided by Peter Klinko and John Hopkins.

\bibliographystyle{apsrev}
\bibliography{qpk}

\end{document}